\documentstyle[prl,aps,preprint]{revtex}

\begin{document}

%\twocolumn[{
\draft
%\preprint{HEP/123-qed}

\title{Frictional Drag Between Coupled 2D Hole Gases in GaAs/AlGaAs Heterostructures}
\author{C. J\"orger, S.J. Cheng, H. Rubel\cite{Newaddr}, W. Dietsche, R. Gerhardts, P. Specht, K. Eberl and K. v. Klitzing}
\address{Max-Planck-Institut f\"ur Festk\"orperforschung, Stuttgart, Germany}

\date{\today}
\maketitle

\begin{abstract}
%\widetext
%\leftskip 54.8 pt
%\rightskip 54.8 pt
We report on the first measurements of the drag effect between coupled 2D-hole gases. 
We investigate the coupling by changing the carrier densities in the quantum wells,
the widths of the barriers between the gases and the perpendicular magnetic field. From the
data we are able to attribute the frictional drag to phonon coupling,
because the non-parabolicity allows to tune the Fermi wavevector
and the Fermi velocity separately and, thereby, to distinguish between phonon-
and plasmon-dominated coupling.

\end{abstract}

%\narrowtext
%{\bf General Introduction}
The coupling mechanisms between two closely spaced two dimensional (2d) charge systems in
semiconductors have found much interest recently. Even if tunneling between the
layers can be neglected, the layers are coupled by the 
electron-electron interactions which leads to a transfer of momenta 
and a  frictional drag between them. The drag force can be measured by passing a drive 
current $I_{drive}$ through one layer and measuring the resulting voltage drop $V_{drag}$ in the 
other one \cite{gram1}. The
coupling strength is usually stated as the transresistivity
$\rho_T = (W/L) (V_{drag}/I_{drive})$ where
$(W/L)$ is the width to length ratio of the sample. The transresistivity has  been derived
theoretically based on RPA and is found to depend on the imaginary parts of the susceptibilities of the two
layers $\chi_1,\chi_2$, on the interlayer interaction $V_{12}$ and on the dielectric constant 
$\epsilon_{12}$ of the combined system \cite{sivan,jauh}:

%\begin{equation}
%\label{gljau}
%\rho_T = \frac{\hbar^2}{8\pi e^2 p_1 p_2 k_B T}\int_0^{\infty}dq q^3 \int
%\frac{d\omega}{2\pi}|\frac{V_{12}(q,\omega)}{\epsilon_{12}}|^2 \\ \times
%\frac{Im\chi_1(q,\omega)Im\chi_2(q,\omega)}{sinh^2(\hbar \omega / 2 k_B T)},
%\end{equation}

\begin{equation}
\label{gljau}
\rho_T \propto \int_0^{\infty}dq q^3 \int
\frac{d\omega}{2\pi}|\frac{V_{12}(q,\omega)}{\epsilon_{12}}|^2 \\ 
\frac{Im\chi_1(q,\omega)Im\chi_2(q,\omega)}{sinh^2(\hbar \omega / 2 k_B T)},
\end{equation}

%where $p_i$ is the 2D charge density in layer i.
Two-dimensional electron gases (2DEGs) in GaAs/AlGaAs
heterostructures in zero magnetic field have been found to couple via Coulomb 
interaction \cite{jauh}, via the excitation of coupled plasmons \cite{flens},
 and via the exchange of phonons with $q \approx 2k_F$ \cite{gram}.
At zero
magnetic fields the Coulomb coupling is  weak compared to the other two
mechanisms. At temperatures exceeding about $0.2 \, T_F$ 
coupled plasmon modes are excited. This is most efficient if the
Fermi velocities $v_F$ of the two  layers coincide. The phonon coupling, on the
other hand, is maximum if the $k_F$ values in the two layers are identical. Very
recently, it has been suggested that coupled electron-phonon modes may exist under certain conditions 
leading to a vanishing of $\epsilon_{12}$
around $q=2k_F$ \cite{bons2}. No evidence for the existence of these coupled modes has been
reported yet.

No experimental work has been published on the frictional drag between two 2d hole gases
(2DHGs) and only preliminary data exist on mixed (2DEG/2DHG) systems \cite{sivan,rub1}. 
In these systems the phonon coupling should be larger because of the larger effective hole masses. 
On the other hand $T_F$ is also smaller for the same reason and it is not clear if phonon or plasmon 
coupling should be expected to dominate at low temperatures. However, hole
systems offer a unique possibility  to discriminate between the different
mechanisms because their  energy dispersion curves are
nonparabolic and, moreover, this nonparabolicity can be tuned by varying the
shape of the quantum wells  by external fields or by doping. Therefore, these systems offer
the possibility to tune both $k_F$ and $v_F$ and to achieve coincidences of either of these
quantitites at
different densities in the two layers. The study of the drag between electron and hole gases will
furthermore give information about the (in)congruences of the Fermi
surfaces which limits the possibility to observe superfluidity in coupled 2DEG/2DHG
systems \cite{conti} by the reduction of the phase space for Cooper-pair-like scattering.

In this Letter we report the first, to our knowledge, measurements of the
frictional drag between two 2DHGs as well as more detailed results on coupled 2DEG/2DHG
systems. In both cases, the frictional drag is measured as function of the charge
densities in the two layers. We also present data on samples having widely varying barrier
thicknesses and we will discuss the effect of temperature and of magnetic field.
We find a very asymmetric behavior with respect to the densities which is in contrast to
the previously studied purely electronic systems, and we present evidence that
the coupling is well described by phonon exchange and not by plasmon interaction.

%{Experimental details}
The coupled 2DHG samples are prepared in two $20\, nm$ thick quantum wells in
GaAs/AlGaAs  heterostructures. Remote doping was achieved using carbon with a spacer layer 
thickness of 
typically $20\, nm$. Six samples are produced with GaAs quantum wells separated by
$\rm Al_{.3}Ga_{.7}As$ barriers with thicknesses varying from $30$ to $190\,
nm$. In some samples the  doping of one well is placed inside the barrier leading to a strong 
asymmetry between the layers. 
The samples are shaped as a Hall bar geometry with $80\,\mu m$ width and $800\,\mu m$ length.  
Ohmic contacts to both layers are made by diffusion of Au and Zn. Separate contacts to the two layers 
are achieved by using the standard selective depletion technique \cite{gram1}. In this case, metallic front gates and
 p-doped buried backgates are used. Two more gates cover the main part of the Hall bar and 
 allow the independent variation of the carrier densities. Typical hole
 mobilities at $4\,K$ are between $40.000\,cm^2/Vs$ and $80.000\, cm^2/Vs$
 which are 
 reasonably good values for hole gases on (001) surfaces. The hole concentration can be varied  
 typically from zero to about $\rm 5 \!\cdot\! 10^{11} \,cm^{-2} $. 
Details of the coupled 2DHG/2DEG system are already described in \cite{rub1}. In this case the 
distance between the two layers is 340 nm. 

The drag measurements are done by passing drive currents of $100\,nA$ at a frequency of about  
$1\,Hz$ through one of the layers and using lock-in
technique to measure the resulting drag voltage in the other layer. The integrity of the signal is
first controlled by checking that all leakage currents are unmeasurably small and
cannot influence the signal.
 Second, the linearity between drag voltage and drive current value is confirmed, and third, the drag
and drive layers are  interchanged and the signals are found to be identical. Measurements are done 
in a standard 
cryostat at temperatures between 1.5 and 10 K and
in magnetic fields up to 11 T.

% {Results in 2DHG/2DHG systems}

First, we present data on the dependence of the transresistivity $\rho_T$ on the hole concentrations. 
The respective densities are determined by Shubnikov-deHaas measurements at 1.5 K . 
In Fig. \ref{fig1} we show $\rho_T$ as a combined grey tone - contour plot at
2.8 K as a function of the upper and lower hole densities for a coupled 2DHG sample with $d=140 \,nm $. 
In comparison to coupled 2DEG structures we find a rather complicated dependence of $\rho_T$ on the 
densities in the two layers. Particularly remarkable is the fact that there is no symmetry in the data if 
the two densities $p_{upper}$ and $p_{lower}$ are interchanged. At small
densities in both  layers ($\le 3\cdot 10^{11} cm^{-2}$) the maximal coupling strength is found along 
a "ridge" 
(marked by heavy dots) running just below the
condition of equal densities in the two layers. At higher densities, however, this ridge splits
into two ones which run nearly vertically and horizontally in the figure (also marked by dots). 
This asymmetry with respect to the densities in the two layers must be
consequence of the asymmetric  doping of the two wells, because it is also seen
in the other asymmetrically doped samples but not in the symetrically doped ones where we find 
a very broad maximum of the coupling centered 
around the line of equal densities. 

% {Rashba effect and analysis of hole-hole coupling}

The asymmetric doping leads to an effective electric field in the respective quantum wells which 
causes the splitting of the highest hole subband into two. This effect has recently been studied in 
detail in just one hole layer \cite{shay}. The 
splitting of the subbands automatically leads to different values of $k_F$ and 
$v_F$ in the two bands, and $v_F$  is no longer proportional to $k_F$ because of the strong 
nonparabolicity of the dispersion curves. 
Both quantities are available from the dispersion curves which we calculate in 
a self-consistent Hartree approximation based on a $ 4\times4\, k\ast p$-method.
Examples are shown in Fig. \ref{fig2} (a) and (b). The $k_F$ of one branch is
identical for the two layers even if their densities are out of balance.
The loci of identical $k_F$ and  of
identical 
$v_F$   are  plotted in  Fig.\ref{fig2} (c) and  (d), respectively. The cross
marks the case of Fig.\ref{fig2} (a) and (b).
Here we use the values obtained for the [110] direction in the plane. 
In the plots of (c) and (d)
the solid  lines correspond to the case where hole subbands with the same quantum number are 
coupled while the dashed lines indicates coupling between bands with different quantum numbers.
A second line of this type lies far outside of the plot range.
 
Comparison of these plots with the data of Fig.\ref{fig1} shows that only the phonon coupling between 
hole bands with identical quantum numbers agrees with the experimental data. In
Fig. \ref{fig1} the calculated 
locus of equal $k_F$ values in the equivalent bands is indicated as a solid line.
If a similar analysis is  made for the [100]-direction, one finds that the loci
for coinciding $k_F$ respectively $v_F$ values do not differ  significantly from
those shown in Fig.\ref{fig1} and Fig.\ref{fig2}. 
We conclude that the phonon exchange is the main source of coupling between the hole layers. 
The fact that only identical branches of the hole dispersion curves couple with each other is most 
likely due to the acoustical anisotropy of the GaAs which cause that only certain phonon modes 
(e.g. longitudinal or transversely polarised ones) couple to either one branch of the dispersion 
curves \cite{rid}.

% {Results in 2DEG/2DHG systems}

Similar experimental data have been obtained in a coupled 2DEG/2DHG system 
with a barrier of 340 nm. The transresistance as function of hole and electron concentration is shown 
in Fig.\ref{fig3}. These data are obtained at 5 K. Similarly to the case of coupled hole gases one finds an 
asymmetric behavior with respect to the two densities, particularly at large densities. We calculate 
again the $k_F$ and $v_F$ values of this 2DHG layers and compare them with $k_{F,e}=\sqrt{2\pi n}$ 
and $v_{F,e}=\hbar k_{F,e}/m^{\ast}$ of the 2DEG. The only satisfactory match with the experimental 
data is obtained using phonon coupling (i.e. matching the $k_F$ values) between
the electrons and only one (the one with the heavier hole  mass) of the hole branches. 
The resulting locus of equal $k_F$ values is plotted in Fig.\ref{fig3} as 
the heavy line. In this case an angular average of the $k_F$ values of the hole
gas is used. 
An interesting result of Fig.\ref{fig3} is that an approximate congruence as it is realised along the ridge
of maximum coupling
between the hole and the electron Fermi surfaces requires in general quite unequal densities 
between the two gases. In contrast to the case of coupled 2d hole gases, the ridge positions of the coupled
 2DEG/2DHG system shifts significantly with increasing temperature because the phase space for scattering processes
changes very differently for the electrons and the holes, respectively \cite{rub1}.

% {distance dependence}
The coupling between two 2d charge gases depends on the distance between the two layers. 
Theoretical studies of the interaction via plasmons predict a strong decrease of the transresistivity 
with distance $\rho_T
\propto d_{eff}^{-3}$ \cite{flens}. Here $d_{eff}$ is the distance between the
center of gravity of the 
wave functions of the respective charge layers. On the other hand, the theory based on the
exchange of coupled phonons predicts a logarithmic decrease with distance \cite{bons2}.
 In Fig.~\ref{fig4} we show data of
${\rho_T/T^2}$ for six coupled 2DHG systems for two different temperatures as
function of $d_{eff}$ at 
matched densities of 
$3\cdot10^{11} cm^{-2}$. There is some scatter in the data which is probably due to different preparation 
conditions of the samples which were fabricated over a time span of more than one year. Nevertheless, 
the comparison of the data with the two predicted thickness dependences shows
clearly that the logarithmic  dependence is a better description of the data. A fit using ${\rho_T/T^2} \propto
ln(l_{ph}/d_{eff})$ with $l_{ph}$ being the mean free path of the phonons, as suggested in \cite{bons2} 
for the case of damped phonons,
gives $l_{ph}\approx 300\, nm$.
This number is small compared to the mean free paths deduced from thermal conductivity or heat 
pulse data for high quality GaAs but agrees quite well with the typical length scale of inhomogeneities 
along MBE grown layers as observed by AFM studies \cite{yoo}.

% {field dependence}
Finally we investigate the dependence of the transresistivity on perpendicular magnetic fields. In earlier 
studies in 2DEG/2DEG systems a dramatic increase of $\rho_T$ was observed except right under the 
quantum Hall effect conditions where $\rho_T$ vanishes \cite{rub2}. A less dramatic increase was 
seen in the 2DHG/2DEG systems \cite{rub3}. The 2DHG/2DHG systems are different because 
quantisation effects are small in most of our magnetic field and temperature regimes. Experimental 
data are shown in Fig.~\ref{fig5} where $\rho_T$ is plotted as function of a perpendicular magnetic 
field. The barrrier 
thickness is $40\,nm$, the carrier-densitiy is $2.5\cdot10^{11} cm^{-2}$ in both wells. At the lowest 
temperature
$\rm$T=1.5 K, $\rho_T$ reflects the Shubnikov-de-Haas-oscillations of $\rho_{xx}$ but the increase of 
the maxima with field is less than with the previously studied systems involving electrons 
\cite{rub2,rub3}. At higher temperatures these quantisation effects disappear, but now 
$\rho_T$ shows a $decrease$ with $increasing$ magnetic field. At even higher
temperatures the $\rho_T$ seems  to become nearly independent of the magnetic
field. Interestingly, this behavior is just the opposite to  the one of $\rho_{xx}$ which $increases$ with
 magnetic field in the same experimental range. 
The general decrease of $\rho_T$ at intermediate temperatures and its flattening at higher temperatures 
is also observed with densities which were not matched between the two layers and with samples having 
wider barriers. This behavior can be qualitatively understood from the behavior of
the susceptibility  $Im\chi(q,\omega)$ at $q\simeq 2k_F$ in magnetic fields. At zero magnetic fields this function 
has 
a strong maximum near $q\simeq 2k_F$ on which most of the analysis of this paper is based. This 
maximum disappears with magnetic field as was shown theoretically by
Glasser \cite{glas}.  Thus, the coupling via phonons with $q\simeq 2k_F$ is weakened in magnetic 
field. At quantising fields which correspond to the usual situation in systems
involving electrons, this 
argument is no longer applicable because other 
types of interactions are dominant \cite{rub3}.

% {conclusions}

In conclusion we report the first data of the frictional drag between coupled 2d hole gases. 
By variation of doping profiles and the application of gate voltages we vary $k_F$ and $v_F$ 
independently from each other and establish that the coupling mechanism is dominated by 
phonon coupling at wavevectors $\simeq 2k_F$. The coupling between  2d electron and 2d 
hole gases can be described within the same model. We find a logarithmic dependence of the
 coupling on the distance between the layers which agrees with a theoretical prediction of 
 coupled phonon-plasmon modes. For the coupled hole gases we find a decrease with magnetic 
 field as long as we are in the classical regime, which is consistent with the
 expected behavior   of the susceptibilities at large wave vectors.

% {acknowledgments}
This work was supported by the BMBF under the grant BM621/4.
 S.J.C gratefully acknowledges financial support by the DAAD.

\begin{figure}
\caption{Contour plot of the transresistivity as function of the carrier-densities in the upper
  ($p_{upper}$) and the lower ($p_{lower}$) 2DHG at T=2.8 K. The barrier-thickness is 140 nm. White
  areas indicate the strongest ($550\, m\Omega _\Box$), the black ones  ($230\, m\Omega _\Box$)
  the weakest coupling. The dotted
  lines run along the "ridges" of maximal $\rho_T$. The heavy lines show the locus of coinciding $k_F$ 
  values in equivalent subbands in the two layers.\label{fig1}}
\end{figure}

\begin{figure}
\caption{Dispersion curves of the lower (a) and the upper (b) 2DHG which can correspond to a pair of equal $k_F$ 
although the densities are different (marked with a cross in (c)). The loci of coinciding $k_F$ and $v_F$ values are shown in (c) and (d), respectively.  Solid and dashed lines indicate coupling  between equivalent and non-equivalent branches of the 
  dispersion curves.
\label{fig2}}
\end{figure}

\begin{figure}
\caption{Transresistivity  in a coupled 2DHG/2DEG system as function of the
respective carrier  densities. The barrier thickness is 340 nm, the temperature at T=5 K. White areas 
correspond to $\rho_T =20\, m\Omega _\Box$, the black ones to  $7\, m\Omega _\Box$ . The solid line 
corresponds to the case of coinciding $k_F$ in the 2DEG and in the heavy hole
band.\label{fig3}} \end{figure}

\begin{figure}
\caption{Transresisistivity  as a function of distance between two
2DHGs for $T=2.8\,K$  and $7\,K$, respectively. The carrier density is  $3\cdot10^{11}cm^{-2}$. 
The full line corresponds to the
  logarithmic dependence expected for phonon coupling. The dashed line is expected for plasmonic
   coupling.\label{fig4}}
\end{figure}

\begin{figure}
\caption{Lower panel:  $\rho_T$ as function of a perpendicular
magnetic field B at  different
 temperatures. The sample contains a  40 nm  barrier and has 
  matched densities of $2.5\cdot10^{11} cm^{-2}$. Upper panel: 
 corresponding  longitudinal resistances of the lower layer.\label{fig5}}
\end{figure}

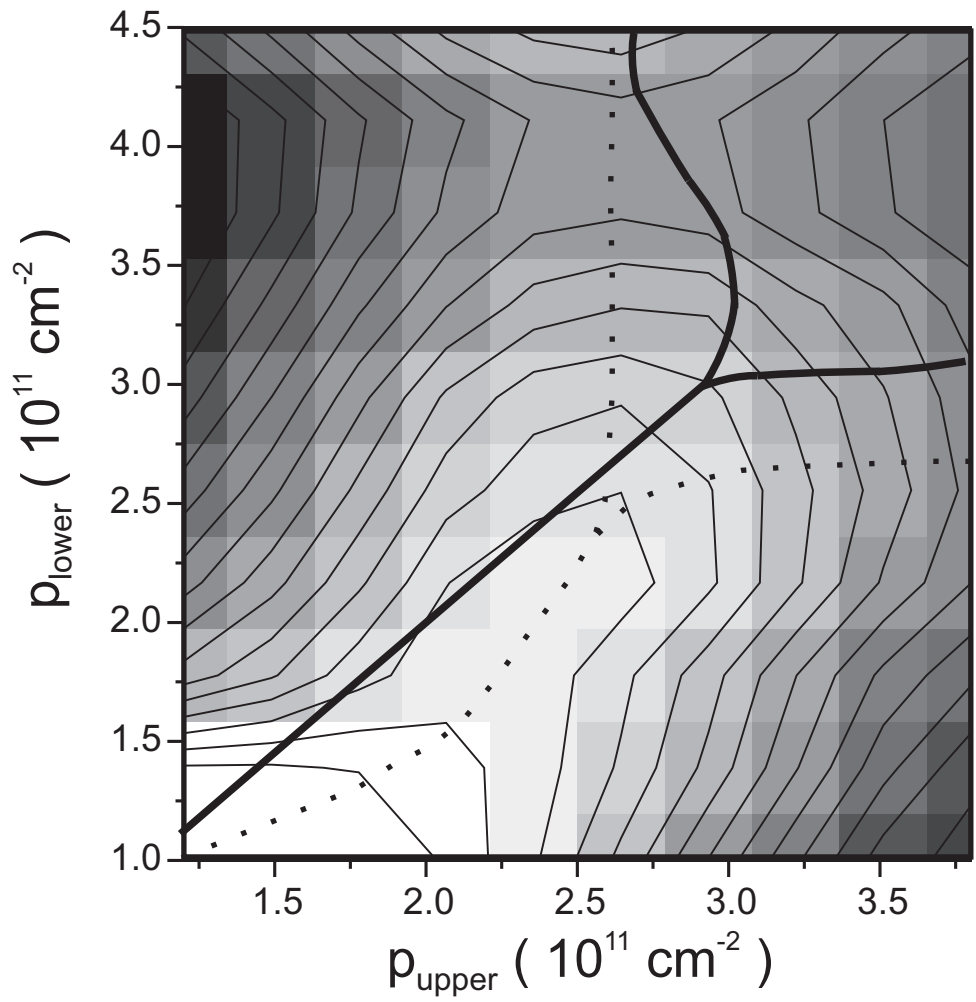
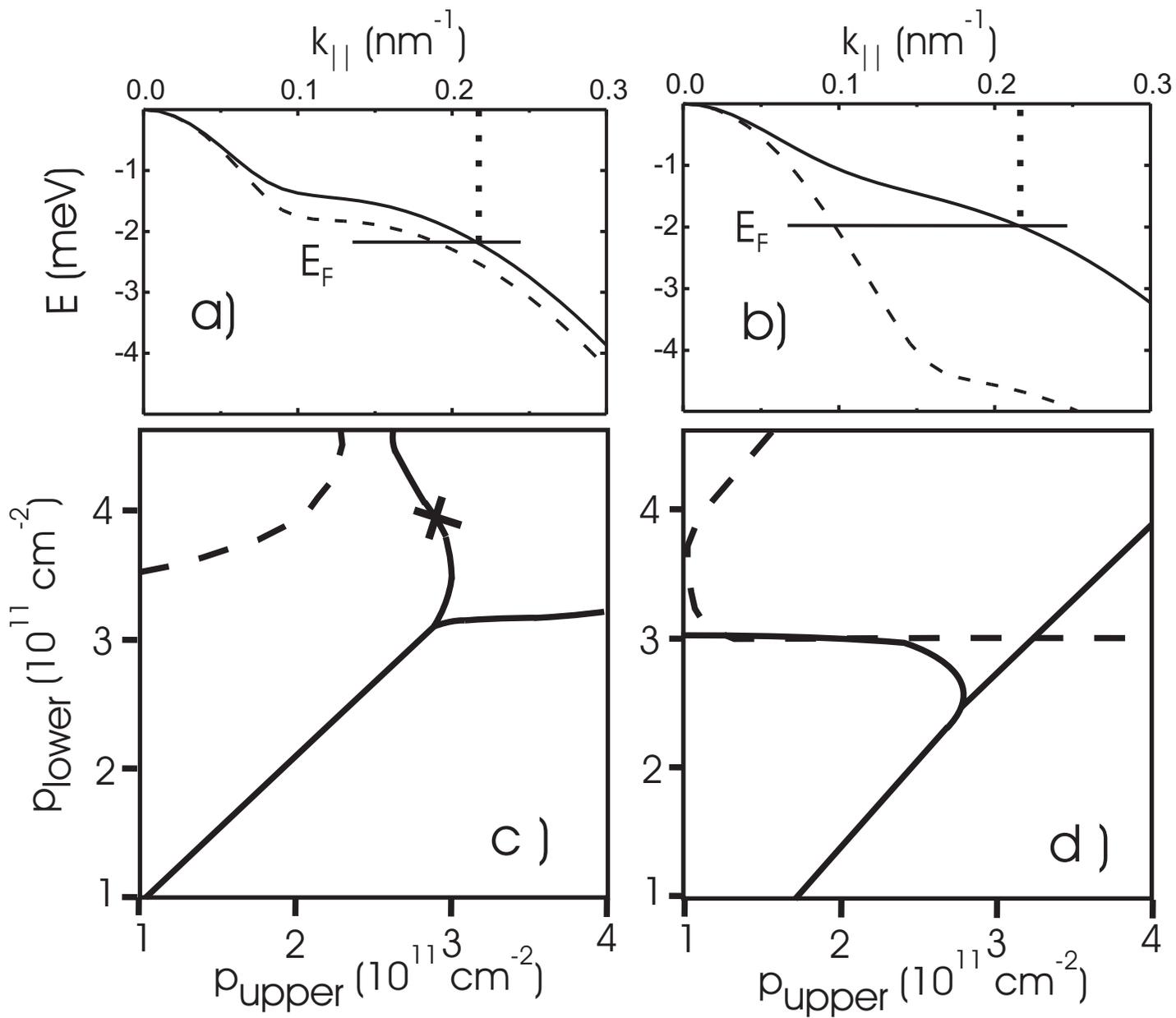
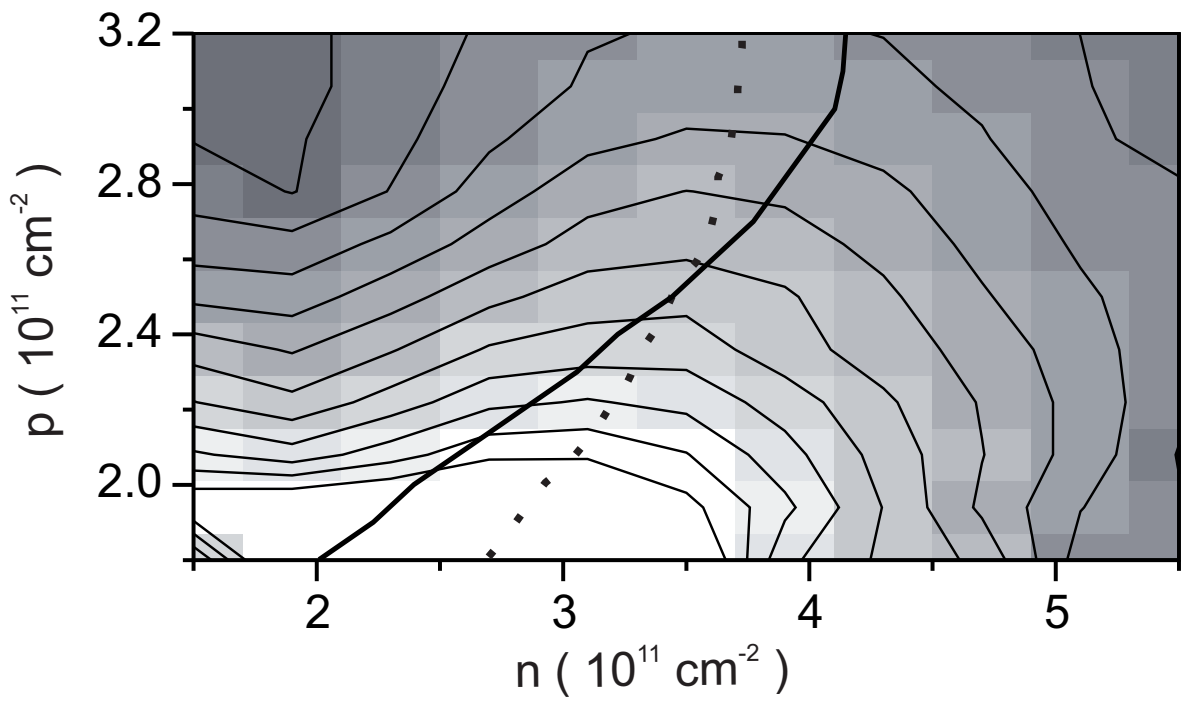
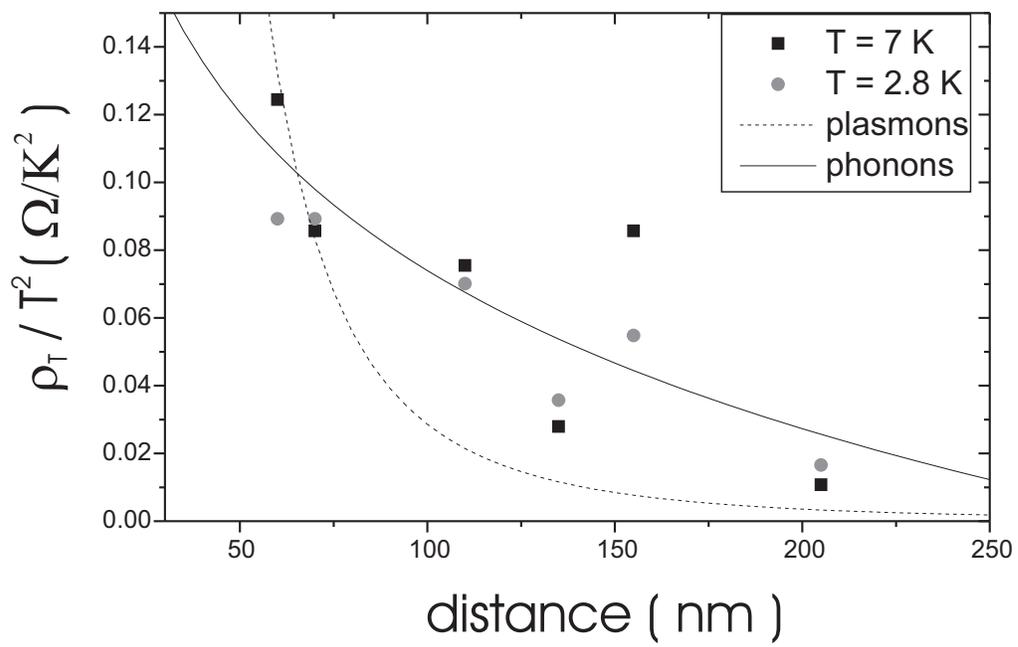
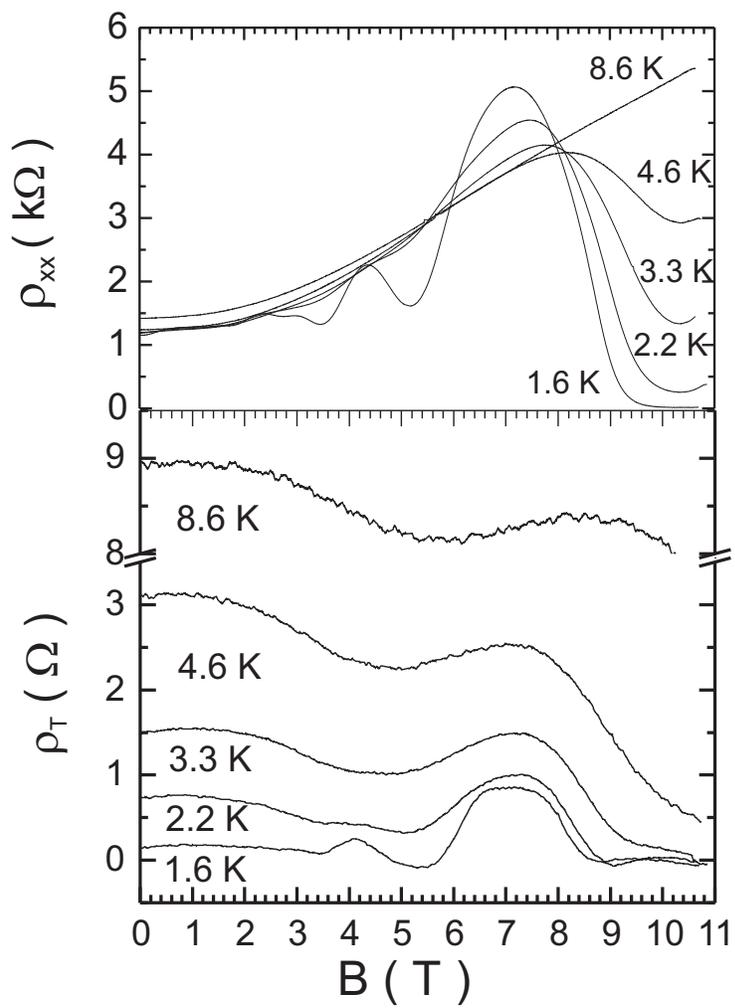
\end{document}